\documentclass[12pt,a4paper,dvips]{article}
\usepackage{a4p}
\usepackage{cite,mcite}
\usepackage{graphicx}
\usepackage{physics}
\usepackage{l3_title,ifthen, Lep}
\journalname{Phys. Lett. B}
 \date{March 13, 2003}
\preprint{2003-012}
\newlength{\capindent}
\newlength{\capwidth}
\newlength{\figwidth}
\newcommand{\icaption}[2][!*!,!]{\hspace*{\capindent}%
  \begin{minipage}{\capwidth}
    \ifthenelse{\equal{#1}{!*!,!}}%
      {\caption{#2}}%
      {\caption[#1]{#2}}
  \end{minipage}}

\def\ee{\mathrm{e^{+}e^{-}}}
\def\WW{\mathrm{W^{+}W^{-}}}

\def\etal{{\it et~al.}}%

\def\rs{\sqrt{s}}

\def\qq{\mathrm{q}\bar{\mathrm{q}}}

\def\WW{\mathrm{W^{+}W^{-}}}

\def\WW{\mathrm{W^{+}W^{-}}}

\begin{document}
\bibliographystyle{l3style}
\begin{titlepage}

\title{ Search for Colour Reconnection Effects in \\
 \boldmath$\ee \rightarrow \WW \rightarrow \mathrm{hadrons}$
 through  Particle-Flow Studies at LEP}

\author{The L3 Collaboration}

%
% The abstract
%
\begin{abstract}
A search for colour reconnection effects in 
 hadronic decays of W pairs is performed with the L3 detector at centre-of-mass energies
 between 189 and 209 \GeV. 
 The analysis is based on 
  the study of the particle flow  between jets associated to the same W boson and between two 
 different W bosons in $\qq\qq$ events.
 The ratio of particle yields in the different  interjet regions
is found to be sensitive to colour reconnection effects implemented in some 
hadronisation models. The data are compared to different models with and
without such effects. An extreme scenario  of colour reconnection is ruled out. 

\end{abstract}
%
% Adds "To be submitted to ..." or "Submitted to ...", if relevant
%
\submitted
% \vspace*{20mm}
% \centerline{To be submitted to Phys. Lett. B. }

\end{titlepage}

%%%%%%%%%%%%%%%%%%%%%%%%%%%%%%%%%%%%%%%%%%%%%%%%%%%%%%%%%%%%%%%%%%%%%%%
\vspace{5cm}

\section{Introduction}

According to the string model of hadronisation, the particles produced 
in the process  $\ee\rightarrow\WW\rightarrow hadrons$
originate, in  the absence of colour reconnection,
 from the fragmentation of two colour singlet strings each of
 which is
stretched between the two quarks from a W boson. 
In this case the hadrons are uniquely associated to a particular W and  
 there is a direct correspondence between the  jets formed by these
hadrons and the primary quarks from the W boson decays. 
 Energy-momentum is separately conserved for each of the W systems.
However,
it has been suggested that interactions may occur between the decay products of
the two W bosons~\cite{gusta,sjost,gh,yellow}.
The main justification for this ``cross-talk'' is the relatively short distance separating the
decay vertices of the W bosons produced  in $\ee$ annihilation
($\approx$ 0.1 fm) compared to the typical hadronic scale (1 fm), 
which implies a large space-time overlap of the two hadronising systems.

The main consequence of these interactions, called Colour Reconnection (CR)
effects, is a modification of the distribution in phase space of hadrons.
CR effects are thought to be suppressed in the
hard perturbative phase, but may be more important in the soft gluon 
emission regime~\cite{sjost}.
 While hard gluons, with energy greater than the W width, are radiated independently
from different colour singlets,
  soft gluons could in principle  be affected by the colour strings of both decaying W's.
Such CR would affect the number of  soft particles in specific
 phase space regions, especially  outside the jet cores.

The study of CR is interesting not only for probing  QCD dynamics but also 
for determining a possible bias in the W mass measurement
in the four-quark channel.
CR could affect the invariant masses of jet pairs originating from
W decays. Therefore
 the precision with which the W mass may be determined using the four-quark channel
depends strongly on the understanding of CR effects.
Events where only one W decays hadronically are unaffected by CR.

Previous LEP  studies of CR, performed at centre-of-mass energy $\rs \le$ 183 \GeV,  
were based on charged particle multiplicity and momentum 
distributions~\cite{lepnchcr}.

The analysis presented in this paper uses the 
method suggested in Reference~\citen{crdom} based on energy and particle flow
to probe the string topology of four-quark events to search for  particular
 effects of particle depletion and  enhancement.
The results are based on 627 pb$^{-1}$ of data collected 
with the L3 detector~\cite{l3detect} at  $\rs$=189$-$209 \GeV.
Comparisons with various  models are made at detector level and the compatibility with
the existence of CR effects in various models is investigated.
 
%%%%%%%%%%%%%%%%%%%%%%%%%%%%%%%%%%%%%%%%%%%%%%%%%%%%%%%%%%%%%%%%%%%%%%%%%%%%%%
\section{Colour Reconnection Models}
Several phenomenological models have been 
proposed~\cite{sjost,gh,geiger,arcr,hwcr,rathsman}
to describe  CR effects in  $\ee\rightarrow\WW\rightarrow hadrons$
 events.
The analysis presented in this paper is performed with  some of those
CR models, which are
implemented in the PYTHIA~\cite{pythia}, ARIADNE~\cite{ariadne}
and HERWIG~\cite{herwig} Monte Carlo (MC)  programs. 

We investigate two models by  Sj\"ostrand and Khoze~\cite{sjost} implemented
in PYTHIA. They are based on
rearrangement of the string configuration during the fragmentation
process. They follow
the space-time evolution of the strings and allow local reconnections
if the strings overlap or cross,
 depending on the string definition (elongated bags  or vortex lines).

In the type I model (SKI)  the strings are associated with colour flux tubes having a
 significant transverse extension. The  reconnection occurs when these tubes 
 overlap and only one reconnection
  is allowed, the one with the largest overlap volume. The reconnection
   probability depends on this  volume of overlap and is controlled by 
   one free parameter, $k_{\rm{I}}$, which can be varied in the model 
to generate event samples  with different fractions of reconnected events.
The relation with the event reconnection probability ($P_{\rm{reco}}$) is given by 
the following formula:
\begin{eqnarray}
P_{\rm{reco}} = 1-{\rm{exp}}(-f k_{\rm{I}})
\end{eqnarray}
where $f$ is a function of the overlap volume of the two strings, which 
depends on W-pair kinematics varying with $\rs$.
The default value of $k_{\rm{I}}$ is 0.6~\cite{sjost}, 
which corresponds
to  a reconnection probability of about 30\% at $\rs$=189 \GeV. 
This analysis is performed with three different values of  
 $k_{\rm{I}}$: 0.6, 3 and 1000, corresponding to reconnection probabilities
 at  $\rs$=189 \GeV\ of about 30\%, 66\% and nearly 100\%, respectively.

 In the type II model (SKII)  the strings have no lateral extent
and  the  reconnection occurs, with unit probability, when they cross. The 
fraction of reconnected events in this model is of the order of 30\%
 at $\rs$=189 \GeV.

 The  CR model  implemented in ARIADNE is based on reconnection
of  coloured dipoles before the string fragmentation takes place~\cite{arcr}.
   In the AR2 scheme, which is investigated here, reconnections are allowed
if they reduce the string length.  While reconnections within a W are
allowed at all scales, those between W's are only allowed after the parton
showers have evolved down to gluon energies less than 2 \GeV.
 At $\rs=$189 \GeV\ they affect about 55\% of the events.

The CR scheme implemented in HERWIG is, as for the string fragmentation, a local
phenomenon since the cluster fragmentation process follows the space-time 
development. In this model~\cite{hwcr} the clusters are rearranged if their space-time
extension is reduced. This rearrangement occurs with a probability equal
to 1/$N^{2}_{\rm{colour}}$, with default value $N_{\rm{colour}}$ = 3, giving
about 23\% of reconnected events.

All probabilities discussed above are derived as fraction of events where at
least one reconnection occurs either within the same W or between two W's.

%%%%%%%%%%%%%%%%%%%%%%%%%%%%%%%%%%%%%%%%%%%%%%%%%%%%%%%%%%%%%%%%%%%%%%%%%%%%%%

\section{Event Selection}

The energy measured 
in the electromagnetic and hadronic calorimeters and in the tracking chamber
is used to select  $\ee\rightarrow\WW\rightarrow hadrons$ events. 
 The total visible energy 
(${E_{\mathrm{vis}}}$) and the energy imbalance parallel (${E_{\parallel}}$)
and perpendicular (${E_{\perp}}$) to the beam direction are measured. 
 The number of clusters, defined as objects obtained from a non-linear 
combination of charged tracks with a transverse momentum greater than 100 \MeV\
and calorimetric clusters 
with a minimum energy of 100 \MeV, is denoted  by 
${N_{\mathrm{cluster}}}$.
The selection criteria are:
\vspace*{-0.1cm}
\begin{center}
   ${E_{\mathrm{vis}}}/ \rs > 0.7$; 
    \hspace*{1cm}  ${E_{\perp} / E_{\mathrm{vis}}}  < 0.2$; 
  \hspace*{1cm}  ${|E_{\parallel}| / E_{\mathrm{vis}}}  < 0.2$;
   \hspace*{1cm}  ${N_{\mathrm{cluster}}} \geq 40 $.
\end{center}   
In addition the events must have 4 jets reconstructed with the Durham algorithm~\cite{kt} 
with $y_{\rm{cut}}$ = 0.01.
To reduce the contamination from
semileptonic W decays, events with energetic  $\mu$ or $\rm{e}$  are rejected.
Events with hard initial state radiation (ISR) are rejected as described in
Reference~\citen{qcdlep2}.
Additional criteria select events with nearly perfect quark-jet association,
necessary for the study of particle and energy flow between jets. 
The two largest interjet angles are required to be between $100^{\circ}$ and $140^{\circ}$ and
 not adjacent.
The two other interjet angles must be less than $100^{\circ}$.
This selection guarantees 
similar sharing of energy between the four primary partons with the two strings 
evolving back-to-back and similar interjet regions between the two W's.
The above cuts are optimized by studying MC $\WW$ events at $\rs$=189
\GeV\ using the KORALW~\cite{koralw} MC generator interfaced with the PYTHIA 
fragmentation model without CR.
Relaxing the angular criteria increases the efficiency 
but gives lower probability to have correct 
 W-jet pairing  due to the more complicated event topology.

The number of selected events, the number of expected events, the selection
efficiency and the percentage of correct pairing are given in
Table~\ref{tab:selflow}.
After applying all the cuts the full sample contains 
666 events with an average efficiency of 12\% and a purity of about 85\%
for $\ee\rightarrow\WW\rightarrow hadrons$.
The average probability to have the correct pairing between the W bosons and their
 associated  jets  is estimated to be 91\%.

The  background is composed of  $\qq (\gamma)$ events
and   Z-pair production events, in similar amounts. Background from semileptonic W pair decays
is found to be negligible (less than 0.3\%).  The $\qq (\gamma)$ 
 process is  modeled with the KK2F MC program~\cite{kk2f}, interfaced
with JETSET~\cite{jetset} routines to describe the QCD processes, and
the background from Z-pair production is simulated with PYTHIA.
  For CR studies W-pair events are simulated with PYTHIA.
All MC samples  are passed through a realistic detector simulation~\cite{l3-simul}
which takes into account  time dependent detector effects and  inefficiencies.

\section{Particle- and Energy-Flow Distributions}
\begin{figure}[htbp]
\begin{center}
    \includegraphics*[width=14cm]{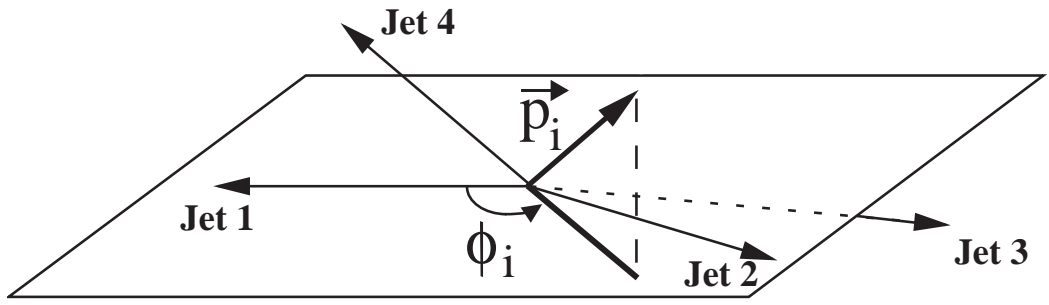}
\end{center}
\caption[]{Determination of the $\phi_{i}$ angle for the particle $i$.}
\label{fig:plane}
\end{figure}

The algorithm  to build the particle- and energy-flow distributions~\cite{crdom} 
(Figure~\ref{fig:plane}) starts by
 defining the plane spanned by the most energetic jet (jet 1) and
the  closest jet making an angle with jet 1 greater than  $100^{\circ}$
 which is most likely associated to the same  W (jet 2).
For each event, the momentum vector  direction of each  particle is then
projected on to this plane.
The particle and energy flows are measured as a function of the angle, $\phi$, 
 between 
jet 1 and the projected momentum vector for the particles located between jets 1 and 2.  
In order to take into account the fact that the W-pair events are not   planar 
a new plane is defined for each remaining pair of adjacent jets.
In this  four-plane configuration the angle $\phi$ is defined as increasing from jet 1 toward jet
2, then to the closest jet from the other W (jet 3) toward the 
remaining jet (jet 4) and back to jet 1. 
The angle $\phi_{j,i}$ of a  particle $i$ having a projected momentum vector located
 between  jets $j$ and $j+1$
is calculated in the plane spanned by these two jets. 
A particle $i$ making an angle $\phi_{i}$ with respect to jet 1 adds an entry equal to 1
in the particle-flow distribution and adds an entry equal to its energy, normalised
 to the total event energy,
in the energy-flow distribution for  the corresponding $\phi$ bin.
  
The distributions are calculated using, for the particle definition,
 the clusters defined in the previous section.

\begin{figure}[htbp]
\begin{center}
    \includegraphics*[width=8.2cm]{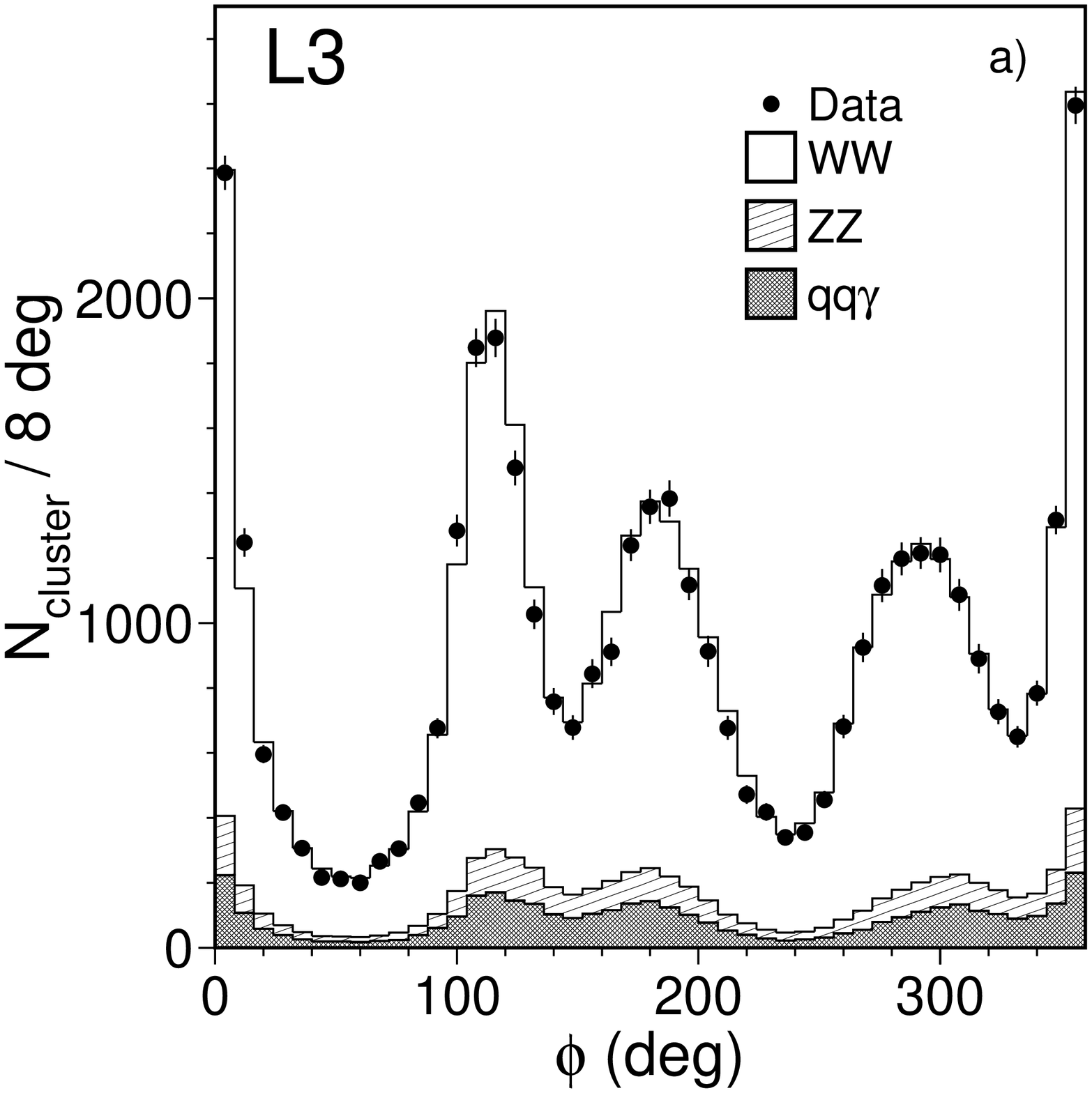}
    \includegraphics*[width=8.2cm]{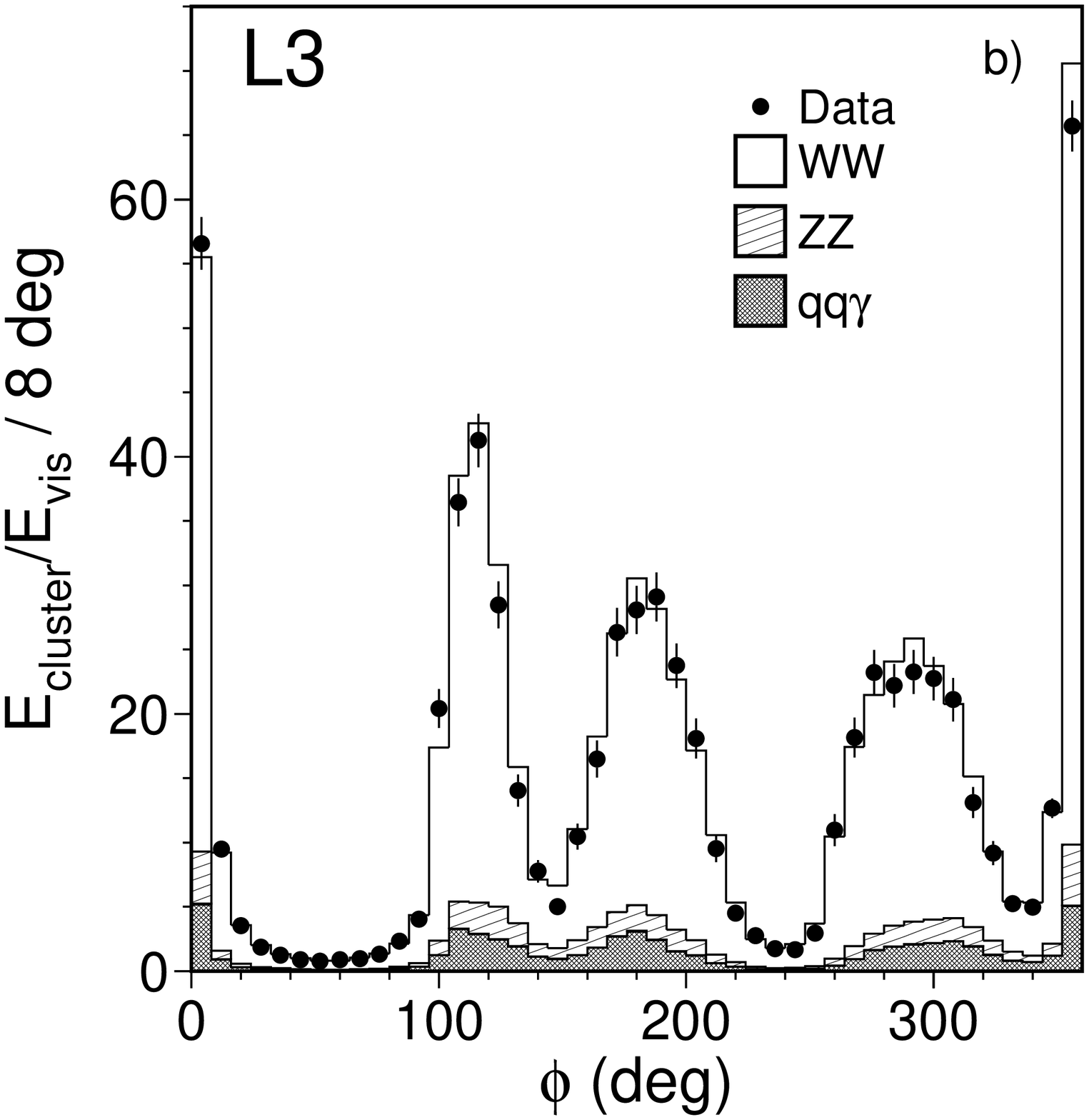}
\end{center}
\caption[]{a) Particle- and b) energy-flow distributions 
at $\sqrt{s} = 189-209 \GeV$ for data and MC predictions.}
\label{fig:flow2}
\end{figure}

 Figure~\ref{fig:flow2} shows the particle- and energy-flow distributions obtained
 for the data and the MC
predictions at detector level by using only the first plane for projecting all the
particles.
The data and MC distributions agree over the full angular range in both cases. 

In order to compare the interjet regions 
the angles in the planes are rescaled by the angle between the two closest jets.
For a particle $i$ located between jets $j$ and $j+1$ the rescaled 
angle is
\begin{eqnarray}
 \phi_{i}^{\mathrm{resc}} = j - 1 +\frac{\phi_{j,i}}{\psi_{j,j+1}}
\end{eqnarray}
 where $\phi_{j,i}$ is the angle between jet $j$ and particle $i$ and  $\psi_{j,j+1}$
 is the angle between jets $j$ and  $j+1$.
 With this definition the four jets have fixed rescaled angle values
 equal to 0, 1, 2 and 3.
 
\begin{figure}[htbp]
\begin{center}
    \includegraphics*[width=12.5cm]{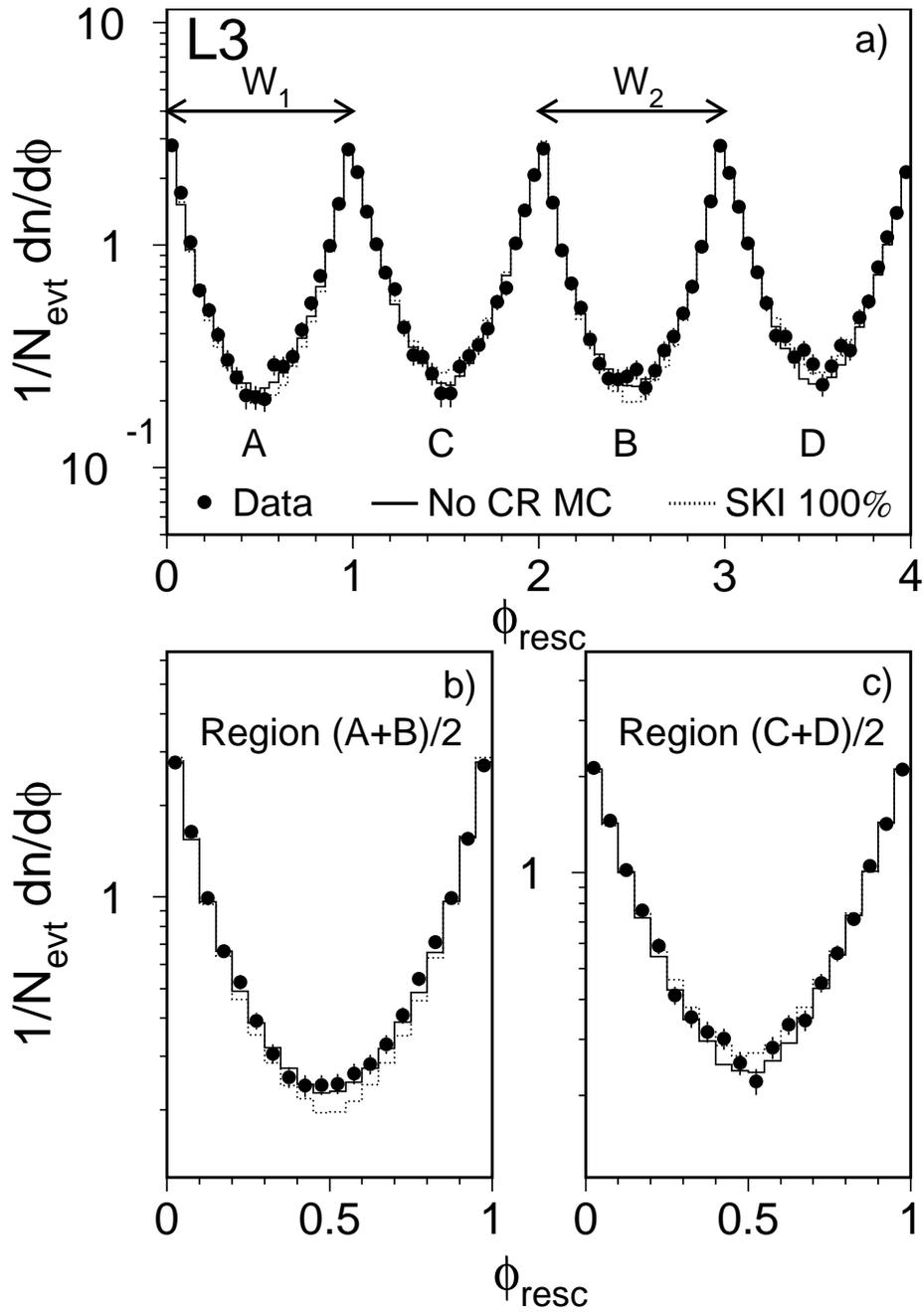}
\end{center}
\vspace*{-1cm}
\caption[]{a) Particle-flow distribution as a function of the rescaled angle
for data and for PYTHIA MC  predictions  without CR,  and
with the SKI 100\% model.  Distributions of b)  combined intra-W particle flow and
c) combined inter-W particle flow.}
\label{fig:flow3}
\end{figure}

Figure~\ref{fig:flow3}a shows the rescaled particle-flow distribution normalised to the
number of events
   after a bin-by-bin  background
subtraction for the data  and  MC predictions 
without CR  and for the SKI model with $k_{\rm{I}}$=1000, later referred to as SKI 100\%.
As expected, the latter shows some depletion in the number of particles  in 
the intra-W regions spanned by the two W bosons (regions A and B)
 and some particle enhancement in the two inter-W regions (regions C and D)
 when compared to the model without CR (no-CR).

%%%%%%%%%%%%%%%%%%%%%%%%%%%%%%%%%%%%%%%%%%%%%%%%%%%%%%%%%%%%%%%%%%%%%%%%%%%%%%
To improve the sensitivity to CR effects the particle flows in  regions A and B 
  are averaged as are the particle flows in regions C and D.
The results are shown in  
Figures~\ref{fig:flow3}b and \ref{fig:flow3}c where the angle is redefined to be in the range [0,1].
MC studies at particle level 
 with particles  having a momentum greater than 100 \MeV\
show that the CR effects are consistent with the detector level results
and have similar magnitudes.

\begin{figure}[htbp]
\begin{center}
    \includegraphics*[width=8.cm]{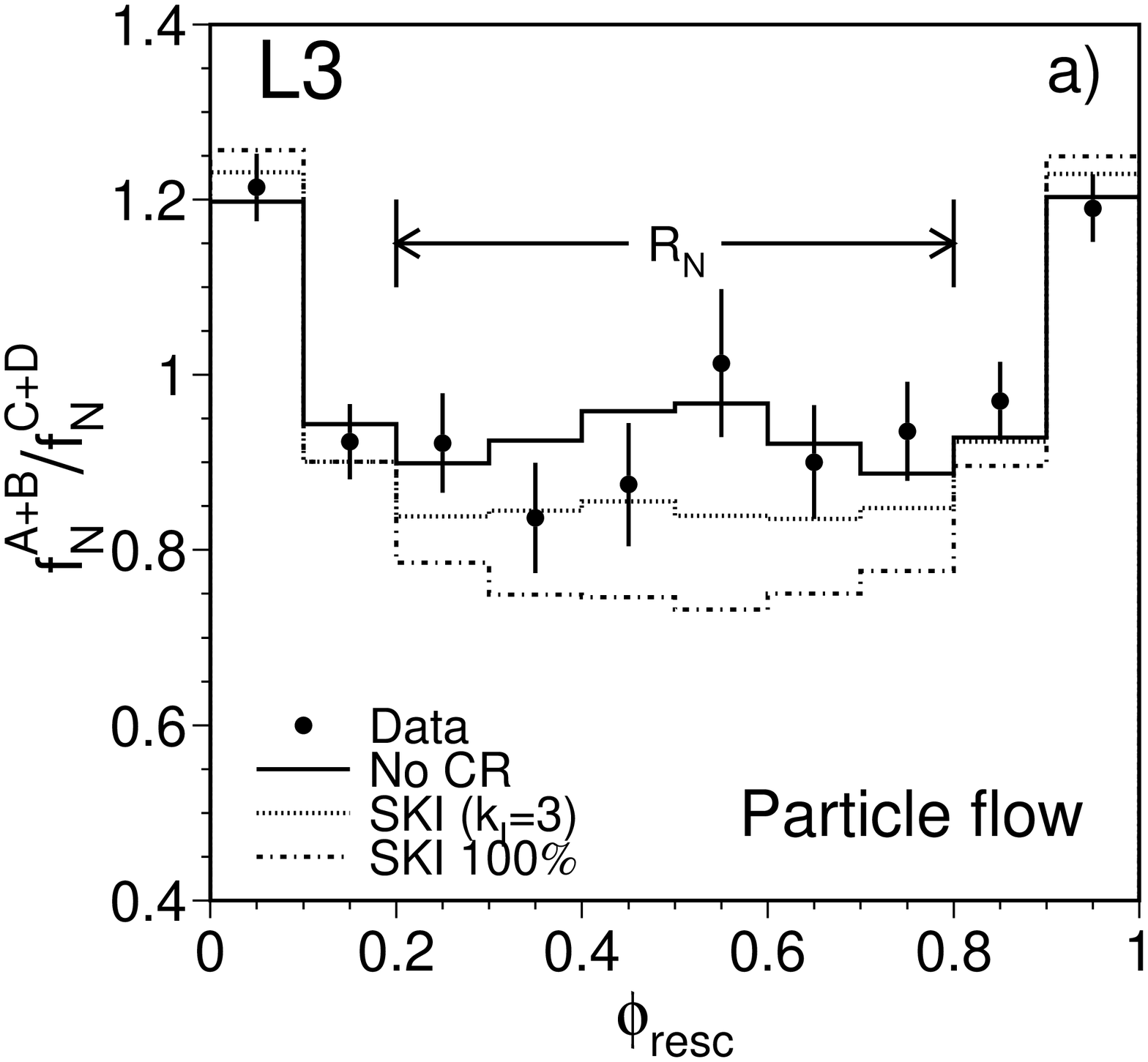}
    \includegraphics*[width=8.cm]{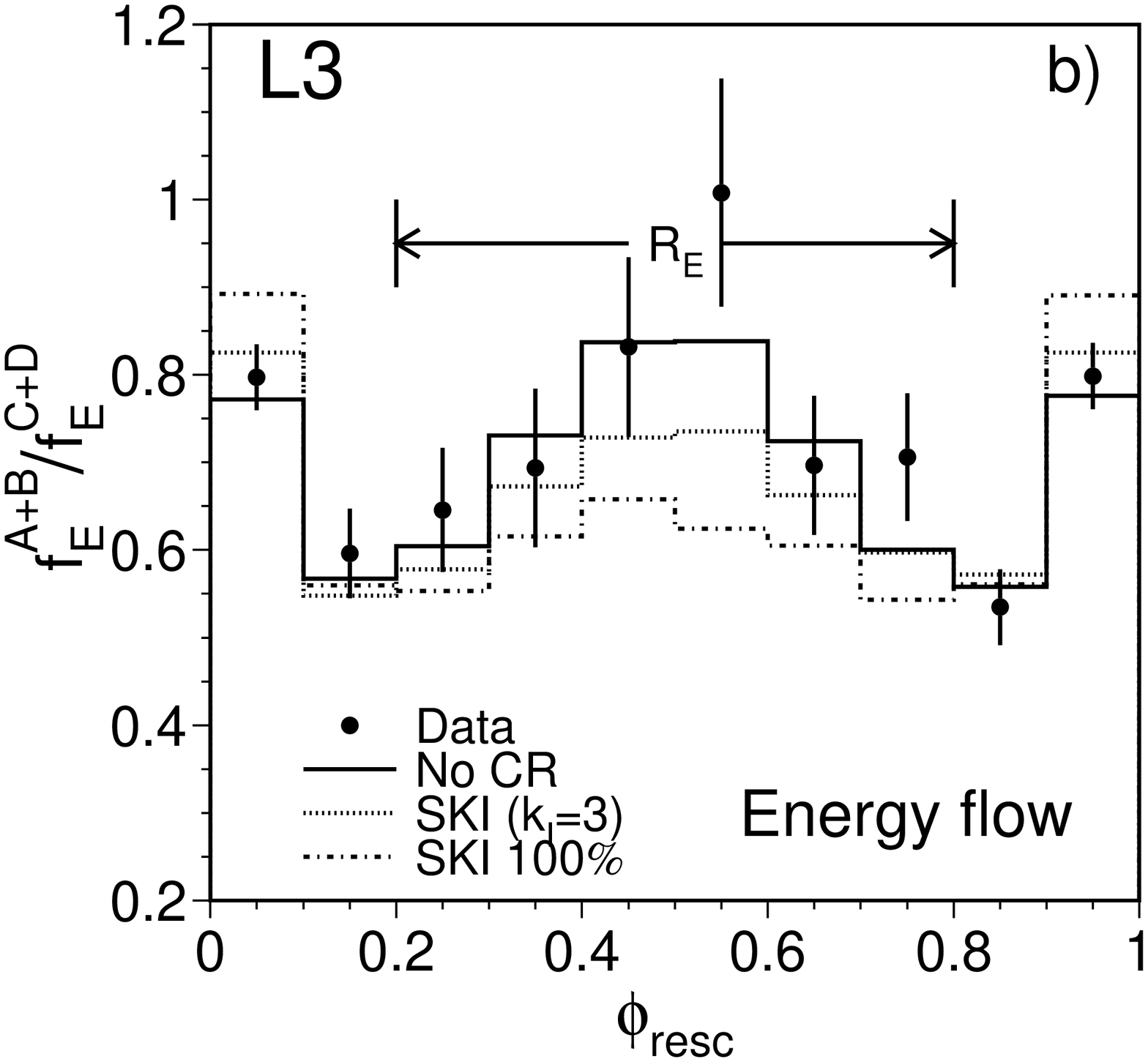}
\end{center}
\caption[]{ Ratio  of a) particle- and b) energy-flow distributions (Equation 4) in regions A+B 
to that in
 regions C+D. Statistical uncertainties are shown.
}
\label{fig:flowrat}
\end{figure}

The ratio of the particle flow between the quarks from the 
same W  to that between quarks from  different W's  
is found to be a sensitive observable to 
cross-talk effects as predicted by the SKI model.
These ratios, computed from the particle- and energy-flow distributions at 
detector level, are shown in Figure~\ref{fig:flowrat} for the data,
 the PYTHIA prediction without CR, the SKI model with 
$k_{\rm{I}}$=3 and SKI 100\%.

The differences between the models with and without CR
are larger in the middle of the interjet regions. 
Therefore, in order to quantify the CR effects the 
ratio $R$ is computed
in an  interval, 0.2  $< \phi_{\mathrm{resc}} <$ 0.8, optimized with respect to the
sensitivity to SKI 100\%. 
The  corresponding variables for
particle and energy flow  are defined as follows:

\begin{eqnarray}
R_{\rm{N}}=  \int_{0.2}^{0.8}{f_{\rm{N}}^{\rm{A+B}}}
\hspace*{0.3cm}{\rm{d}}\phi \left/\ \int_{0.2}^{0.8}{f_{\rm{N}}^{\rm{C+D}}}
\hspace*{0.3cm}{\rm{d}}\phi 
\hspace*{0.3cm}\mathrm{ and}\hspace*{0.3cm}
R_{\rm{E}}=  \int_{0.2}^{0.8}{f_{\rm{E}}^{\rm{A+B}}}
\hspace*{0.3cm}{\rm{d}}\phi \right/\ \int_{0.2}^{0.8}{f_{\rm{E}}^{\rm{C+D}}}
\hspace*{0.3cm}{\rm{d}}\phi 
\end{eqnarray}
where, in a region $i$,
\begin{eqnarray}
f_{\rm{N}}^{i}= \frac{1}{N_{\rm{evt}}}\frac{{\rm{d}}n}{{\rm{d}}\phi}
\hspace*{0.3cm}\mathrm{ and}\hspace*{0.3cm}
f_{\rm{E}}^{i} = \frac{1}{E}\frac{{\rm{d}}E}{{\rm{d}}\phi}
\end{eqnarray}

The measured values of $R_{\mathrm{N}}$ and $R_{\mathrm{E}}$ obtained at each centre-of-mass 
energy are summarised in
Table~\ref{tab:result}. Correlations
in the particle rates between the four interjet regions are
taken into account by constructing the full covariance matrix.
 This results in an increase of about 20\% of the statistical uncertainty.
The  values obtained with the complete data sample are:
\begin{eqnarray*}
R_{\rm{N}}  =  0.911 \pm 0.023\hspace*{0.1cm}(\rm{stat.})\\
R_{\rm{E}}  =  0.719 \pm 0.035\hspace*{0.1cm}(\rm{stat.}) 
\end{eqnarray*}
An estimate of the sensitivity to the SKI 100\% model, shows that  $R_{\rm{N}}$
 is 2.6 times more sensitive than $R_{\rm{E}}$. 
Accordingly, the following results and discussion are only based on $R_{\rm{N}}$.

\begin{figure}[htbp]
\begin{center}
    \includegraphics*[width=12.cm]{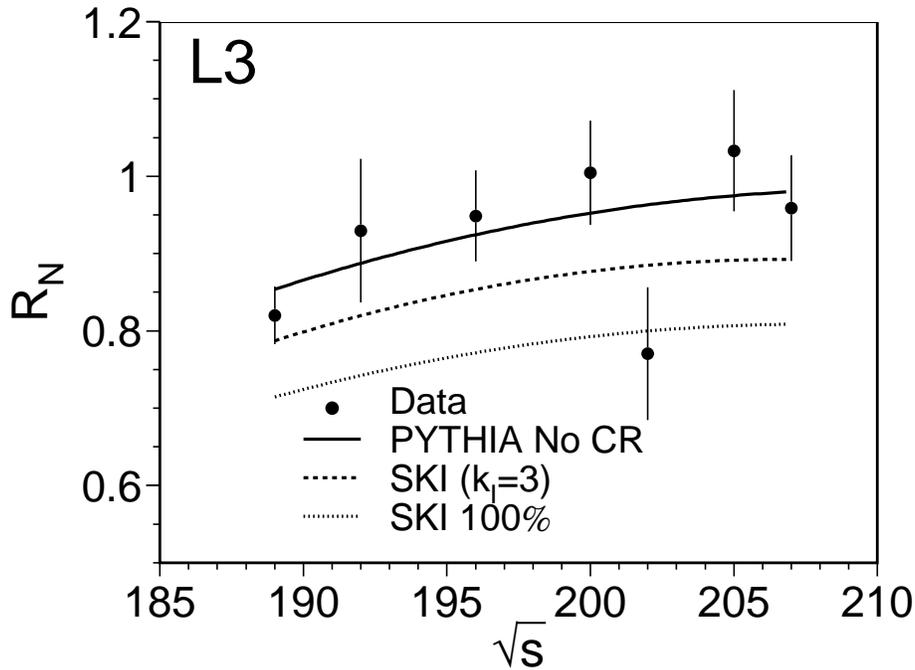}
\end{center}
\caption[]{The ratio $R_{\rm{N}}$ as a function of $\rs$ at detector level
 for data and  PYTHIA no-CR and  SKI model predictions.
The parametrisation of the energy dependence is obtained by  fitting  a second order
 polynomial function to
the predicted MC dependence.
The parametrisation  obtained with PYTHIA no-CR gives
$R_{\rm{N}}(\rs)/R_{\rm{N}}$(189 {\rm{GeV}}) = $-3.07 \times 10^{-4} s + 0.1297 \rs -12.56$. 
The dependence obtained with the SKI model ($k_{\rm{I}}$= 3) leads to a 
2.3\% change in the average rescaled $R_{\rm{N}}$ value at 189 \GeV.
}
 
\label{fig:ratio}
\end{figure}

Figure~\ref{fig:ratio} shows the measured $R_{\rm{N}}$ as a function of
$\rs$ together  with PYTHIA no-CR and SKI model predictions.
 The energy dependence originating from  the 
different pairing  purities 
and jet configurations is in agreement with the model predictions.
For the PYTHIA SKI predictions,
the ratio decreases with the reconnection probability over the whole 
energy range with similar magnitude. 
The data indicate little or no CR.

\section{Semileptonic Decays}
To verify the quality of the MC simulation of the $\mathrm{W}\rightarrow\qq$ 
fragmentation process and the possible biases
 which may arise when determining the particle yields between reconstructed jets in the detector, 
 the particle- and energy-flow distributions are investigated in
 $\ee\rightarrow\WW\rightarrow\qq l\nu$ where $l=\mathrm{e},\mu$.
For this analysis events are selected with high multiplicity,
large missing momentum and a high energy electron or muon.
The missing momentum is considered as a fictitious particle in order to apply the Durham
 jet algorithm to select  4-jet events with  $y_{\rm{cut}}$=0.01.

\begin{figure}[htbp]
\begin{center}
    \includegraphics*[width=8.2cm]{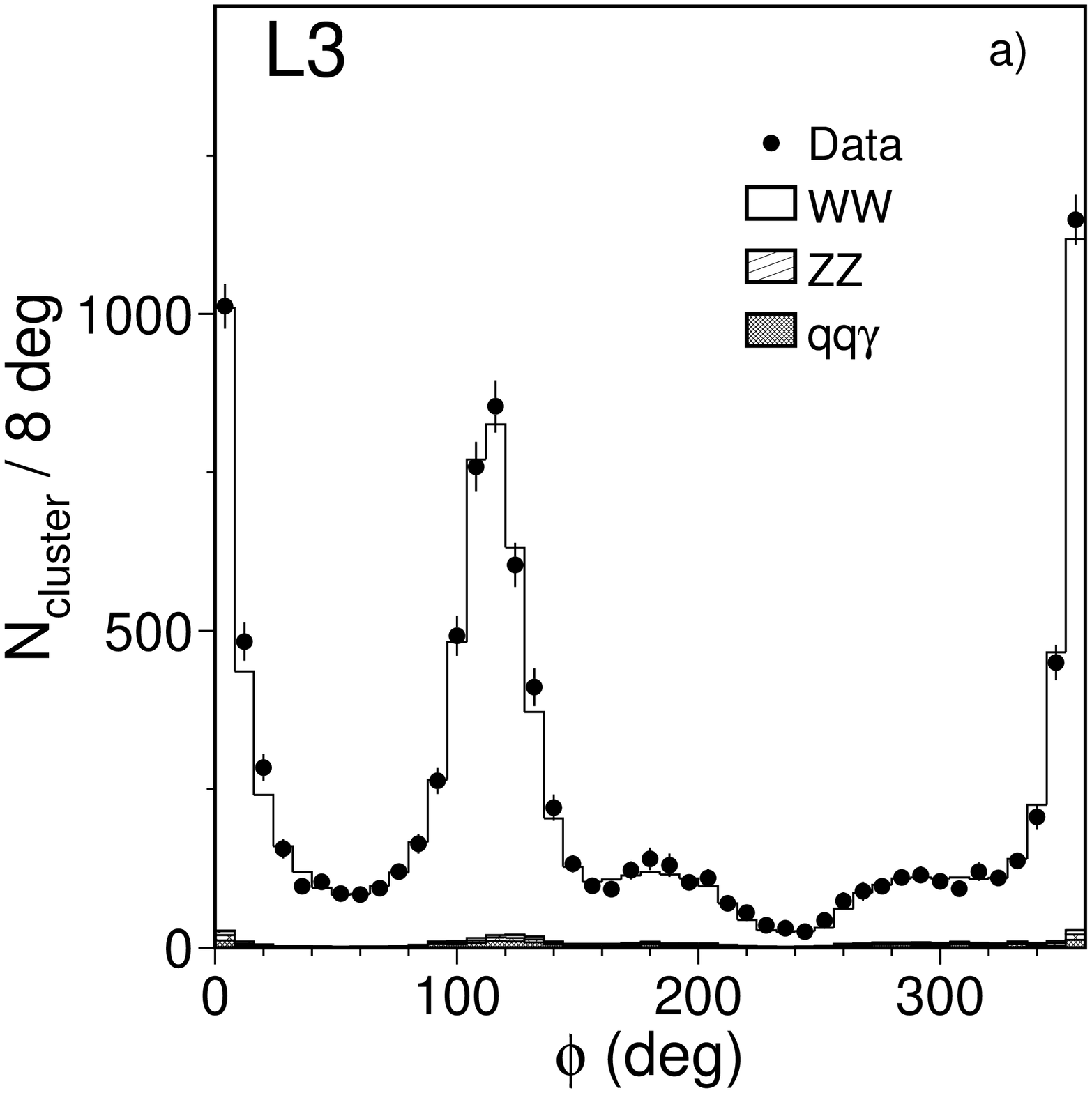}
    \includegraphics*[width=8.2cm]{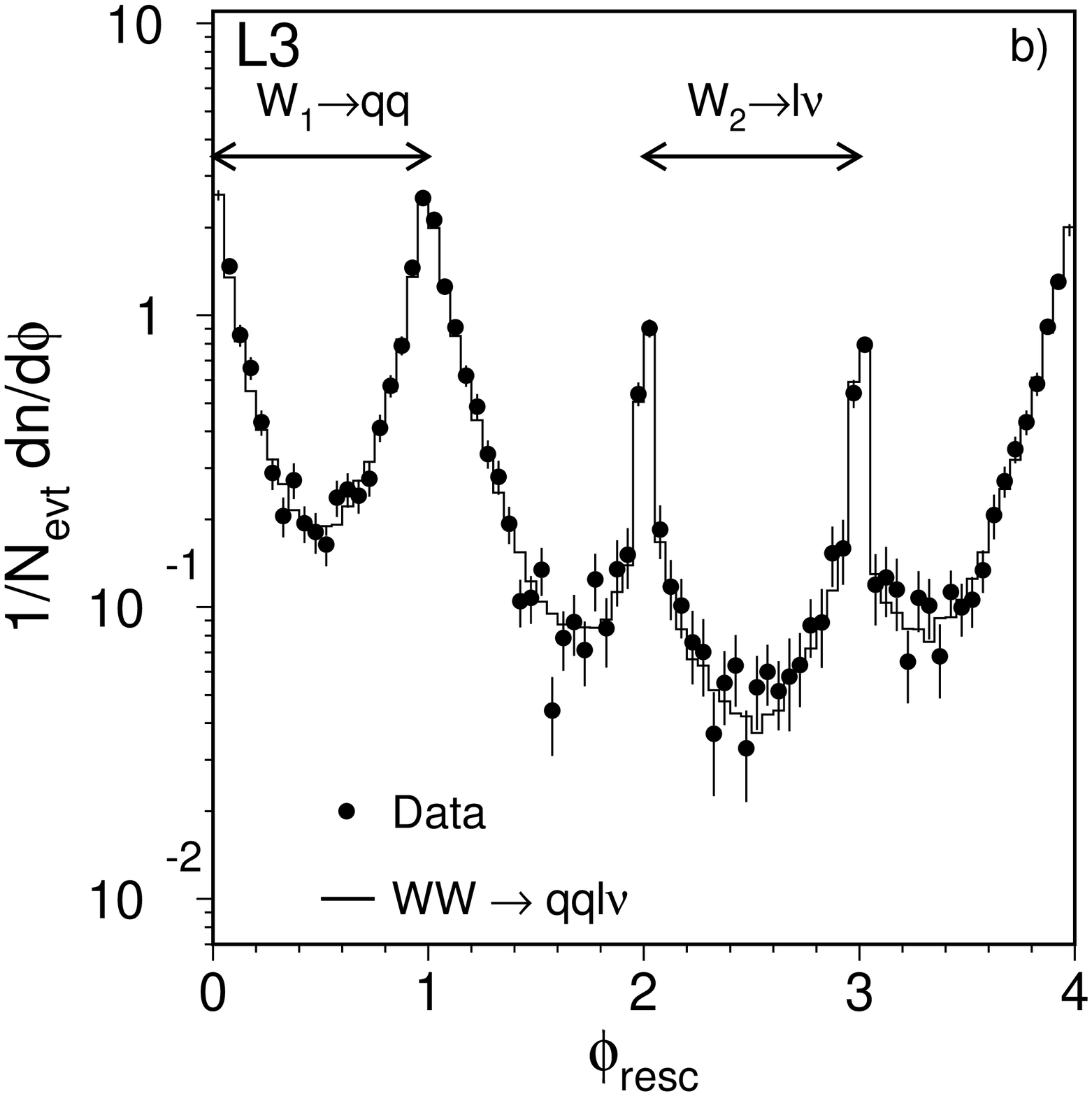}
\end{center}
\vspace*{-1.0cm}
\caption[]{Particle-flow distributions a) before and b) after angle rescaling 
for the semileptonic W decays for 
data and KORALW prediction.}
\label{fig:flowlept}
\end{figure}

\begin{figure}[htbp]
\begin{center}
    \includegraphics*[width=8.cm]{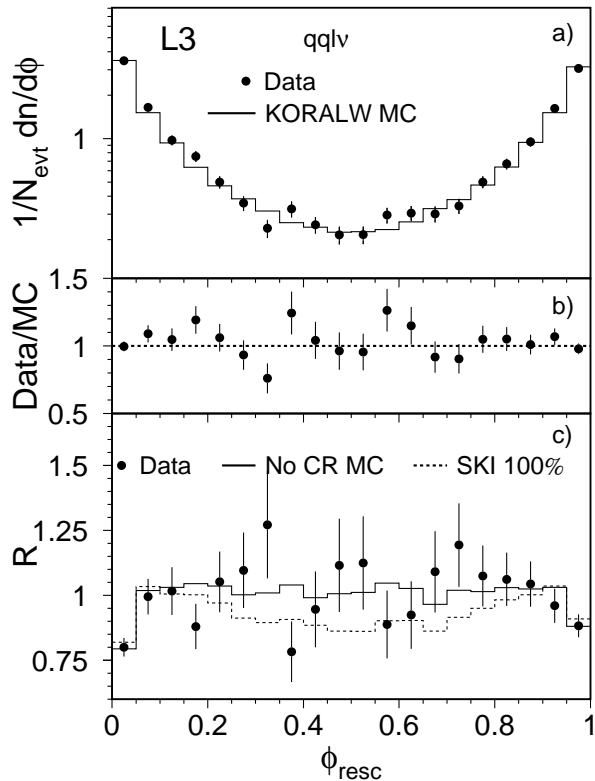}
\end{center}
\vspace*{-1.0cm}
\caption[]{a) Particle-flow distributions as a function of the rescaled angle
for the semileptonic W decays for 
data and the KORALW prediction.
b) Ratio of data and MC as a function of the rescaled angle. c) 
Ratio $R$  of the particle flow in hadronic events
 divided by twice the particle flow in semileptonic events.}
\label{fig:flowlept2}
\end{figure}

  The same angular criteria  on the four interjet angles 
 as applied in the fully hadronic channel are used here. 
The purity obtained after selection is about 96\% and the efficiency 
 is about 12\%. The number of selected semileptonic
events  is 315 with an expectation of 314.5 events.
Particle- and energy-flow distributions are built in a similar way as
 in the fully hadronic channel with the additional requirement that 
 the charged lepton should be in  jet 3 or 4. 
Figure~\ref{fig:flowlept}a
shows the corresponding particle-flow distribution projected on to the plane of 
jets 1 and 2 for  the data 
and the KORALW MC prediction. There is good agreement between data and MC
over the whole distribution. 
Figure~\ref{fig:flowlept}b shows the rescaled particle-flow distribution where 
the  structure of 
the two different W's is clearly visible.
The region between jet 1 and jet 2 corresponds to the hadronically decaying W ($W_{1}$) and
the region between jet 3 and jet 4 corresponds to the W decaying semileptonically ($W_{2}$).
 The  activity in the $W_{2}$ region
is mainly due to low energy fragments from the hadronic decay of the first W.
%%%%%%%%%%%%%%%%%%%%%%%%%%%%%%%%%%%%%%%%%%%%%%%%%%%%%%%%%%%%%%%%%%%%%%%%%%%%%%
A comparison of data and MC for the particle flow obtained by summing the regions 
 $W_{1}$ and $W_{2}$  is shown in Figure~\ref{fig:flowlept2}a.
The ratio between the data and the MC distributions is  shown in Figure~\ref{fig:flowlept2}b.
This ratio is consistent with unity over the whole range.
%%%%%%%%%%%%%%%%%%%%%%%%%%%%%%%%%%%%%%%%%%%%%%%%%%%%%%%%%%%%%%%%%%%%%%%%%%%%%%
This result gives additional confidence in the correctness of the
 modelling of the fragmentation process of  quark pairs according to the 
 fragmentation parameters used in  KORALW and PYTHIA 
 as well as the particle flow definition and reconstruction.

In the absence of CR effects, the  activity found in regions A+B of  a 
fully hadronic event should be equivalent to  twice the particle activity in the 
regions  $W_{1}$+$W_{2}$  of the distribution for a semileptonic event. Figure~\ref{fig:flowlept2}c
shows the ratio of the particle flow in four-quark events divided by twice the particle flow 
in semileptonic events for the data  and the predictions from no-CR PYTHIA MC  and
 the SKI 100\% model. The CR model shows the expected deficit in the hadronic
 channel compared to the semileptonic one. The data are consistent with the no-CR scheme but the 
large statistical uncertainty prevents a quantitative
 statement based on this model-independent comparison. 

%%%%%%%%%%%%%%%%%%%%%%%%%%%%%%%%%%%%%%%%%%%%%%%%%%%%%%%%%%%%%%%%%%%%%%%%%%%%%%
\section{Systematic Uncertainties}

Several sources of systematic uncertainties are investigated. 
The first  important  test is whether the
result depends on  the definition of the particles. The analysis is 
repeated using calorimetric clusters only.
 Half  the difference between the two analyses is assigned as the uncertainty due
to this effect. This is found to be the dominant 
systematic uncertainty.

The second source of systematic uncertainty 
is the limited knowledge of  quark fragmentation modelling.
The systematic effect in the $\qq (\gamma)$ background 
 is estimated by comparing results using 
the JETSET and HERWIG MC programs.  The corresponding uncertainty is
assigned to be half  the  difference between the two models.

The systematic uncertainty from quark fragmentation modelling in W-pair events
is estimated by comparing results using
PYTHIA, HERWIG and ARIADNE MC samples without CR. The uncertainty is assigned as
the RMS  between the $R_{\mathrm{N}}$ values obtained with the three fragmentation models.
Such comparisons between different models  test also possible effects
of different fragmentation schemes
which are not taken into account when varying only fragmentation parameters
within one particular model.
 
Another source associated with fragmentation modelling  is
the  effect of   Bose-Einstein correlations (BEC) in hadronic
W decays. This effect  is estimated by repeating the analysis using
a MC sample with BEC only between particles originating
from the same W. An uncertainty is assigned equal to half the
difference with the default MC which includes full BEC simulation (BE32 option)~\cite{be32} 
in W pairs. 
The sensitivity of the $R_{\rm{N}}$ variable to BEC is found to be small.

The third main source of systematic uncertainty is the background estimation.
The  $\qq (\gamma)$ background which is subtracted corresponds mainly to QCD four-jet 
events for which the rate is not well modelled by parton shower programs. PYTHIA underestimates,
by about 10\%, the four-jet rate in the selected phase space region~\cite{4jetlep}.
A systematic uncertainty is
 estimated by varying the  $\qq (\gamma)$  cross section by $\pm$ 5\% after correcting
 the corresponding background by +5\%.
  This correction increases the value of $R_{\rm{N}}$
  by 0.004.

A last and small systematic uncertainty is associated with  Z-pair production. 
It is estimated by varying the corresponding cross section by $\pm$ 10\%.
This variation takes into account all possible uncertainties pertaining to
the hadronic channel, from final state interaction effects to the theoretical 
knowledge of the hadronic cross section.

A summary of the different contributions to the systematic uncertainty is given
in Table~\ref{tab:syst}.

The ratio   obtained by taking into account the systematic uncertainties  is then:
\begin{eqnarray*}
R_{\rm{N}}  =  0.915 \pm 0.023 \hspace*{0.1cm} (\rm{stat.}) \pm 0.021 \hspace*{0.1cm}(\rm{syst.})
\end{eqnarray*}

\section{Comparison with Models}

 The $R_{\rm{N}}$ values predicted by the PYTHIA no-CR, SKI,  SKII, ARIADNE no-CR, 
AR2, HERWIG no-CR and HERWIG CR  models are given in Table~\ref{tab:model}.

  The data disfavour extreme scenarios of CR.
 A comparison with ARIADNE and HERWIG  shows that
 the CR schemes implemented in these two models 
 do not modify significantly the interjet particle activity in the hadronic W-pair decay events.
  Thus it is not possible to constrain either of these models in the present analysis.

The dependence of $R_{\rm{N}}$ 
 on the reconnection probability is investigated with the SKI model.
For this, four MC samples are used: the no-CR sample and those 
 with $k_{\rm{I}}$=0.6, 3 and 1000. 
In the SKI model the fraction of reconnected events is controlled by the $k_{\rm{I}}$
parameter and the dependence of $R_{\rm{N}}$ on $k_{\rm{I}}$ is parametrized as
$R_{\rm{N}}(k_{\rm{I}}) = p_{1} (1-{\rm{exp}}(-p_{2} k_{\rm{I}}))+p_{3}$ where
 $p_{i}$ are   free parameters.
 A $\chi^{2}$ fit to the data is performed.
The $\chi^{2}$ minimum is at $k_{\rm{I}}=0.08$.  This value corresponds to about 6\%
reconnection probability  at $\rs$=189 \GeV. Within the
 large uncertainty  the result is also consistent with no CR effect. 

The upper limits  on $k_{\rm{I}}$ at 68\% and 95\% confidence level are derived
as 1.1 and 2.1 respectively.
The corresponding reconnection probabilities at $\rs$ = 189 \GeV\ are 45\% and 64\%.
The extreme SKI scenario, in which CR
occurs in essentially all events, is disfavoured by 4.9 $\sigma$.

%%%%%%%%%%%%%%%%%%%%%%%%%%%%%%%%%%%%%%%%%%%%%%%%%%%%%%%%%%%%%%%%%%%%%%%%%%%%%%%
%
%\clearpage
%%%%%%%%%%%%%%%%%%%%%%%%%%%%%%%%%%%%%%%%%%%%%%%%%%%%%%%%%%%%%%%%%%%%%%%%%%%%%%%
% Bibliography
%%%%%%%%%%%%%%%%%%%%%%%%%%%%%%%%%%%%%%%%%%%%%%%%%%%%%%%%%%%%%%%%%%%%%%%%%%%%%%
%
% Style file to use with mcite.
% Use l3style with just cite.
% \bibliographystyle{/l3/paper/biblio/l3stylem}
% \bibliography{l3at183}

%%%%%%%%%%%%%%%%%%%%%%%%%%%%%%%%%%%%%%%%%%%%%%%%%%%%%%%%%%%%%%%%%%%%%%%%%%%%%%%
% The author list
%%%%%%%%%%%%%%%%%%%%%%%%%%%%%%%%%%%%%%%%%%%%%%%%%%%%%%%%%%%%%%%%%%%%%%%%%%%%%%%
%
\newpage
%\section*{Author List}
\typeout{   }     
\typeout{Using author list for paper 261 -  }
\typeout{$Modified: Jul 15 2001 by smele $}
\typeout{!!!!  This should only be used with document option a4p!!!!}
\typeout{   }
%
%
%
%  L A T E X  version!!
%
%
% Make sure that the Lep package has been used!
%\input{Lep.sty}%
%
%\ifx\LepCalled\undefined%
%\typeout{     }%
%\typeout{!!!!!!!!!!!!!!!!!!!!!!!!!!!!!!!!!!!!!!!!!!!!!!!!!!!!!!!!!!!}%
%\typeout{Yikes.  You haven't used the Lep package!}%
%\typeout{Please put \protect\usepackage\protect{Lep\protect} in your preamble,
%         followed by}%
%\typeout{\protect\Lep\protect{1\protect} or \protect\Lep\protect{2\protect}}%
%\typeout{     }%
%\typeout{For now you will get a Lep phase 2 authorlist (may not be right!).}%
%\typeout{!!!!!!!!!!!!!!!!!!!!!!!!!!!!!!!!!!!!!!!!!!!!!!!!!!!!!!!!!!!}%
%\typeout{     }%
%\Lep{2}\fi%

\newcount\tutecount  \tutecount=0
\def\tutenum#1{\global\advance\tutecount by 1 \xdef#1{\the\tutecount}}
\def\tute#1{$^{#1}$}
\tutenum\aachen            % 1
\tutenum\nikhef            % 2
\tutenum\mich              % 3
\tutenum\lapp              % 4
\tutenum\basel             % 5
\tutenum\lsu               % 6
\tutenum\beijing           % 7
\tutenum\bologna           % 8
\tutenum\tata              % 9 
\tutenum\ne                % 10
\tutenum\bucharest         % 11
\tutenum\budapest          % 12
\tutenum\mit               % 13
\tutenum\panjab            % 14 
\tutenum\debrecen          % 15
\tutenum\dublin            % 16
\tutenum\florence          % 17
\tutenum\cern              % 18
\tutenum\wl                % 19
\tutenum\geneva            % 20
\tutenum\hefei             % 21
\tutenum\lausanne          % 22
\tutenum\lyon              % 23
\tutenum\madrid            % 24
\tutenum\florida           % 25
\tutenum\milan             % 26
\tutenum\moscow            % 27
\tutenum\naples            % 29
\tutenum\cyprus            % 30
\tutenum\nymegen           % 31
\tutenum\caltech           % 32
\tutenum\perugia           % 33
\tutenum\peters            % 34
\tutenum\cmu               % 35
\tutenum\potenza           % 36
\tutenum\prince            % 37
\tutenum\riverside         % 38
\tutenum\rome              % 39
\tutenum\salerno           % 40
\tutenum\ucsd              % 41
\tutenum\sofia             % 42
\tutenum\korea             % 43
\tutenum\purdue            % 44
\tutenum\psinst            % 45
\tutenum\zeuthen           % 46
\tutenum\eth               % 47
\tutenum\hamburg           % 48
\tutenum\taiwan            % 49
\tutenum\tsinghua          % 50

{
\parskip=0pt
\noindent
{\bf The L3 Collaboration:}
\ifx\selectfont\undefined%  old style font selection
 \baselineskip=10.8pt
 \baselineskip\baselinestretch\baselineskip
 \normalbaselineskip\baselineskip
 \ixpt
\else%                      new style font selection
 \fontsize{9}{10.8pt}\selectfont
\fi
\medskip
\tolerance=10000
\hbadness=5000
\raggedright
\hsize=162truemm\hoffset=0mm
\def\r{\rlap,}
\noindent

P.Achard\r\tute\geneva\ 
O.Adriani\r\tute{\florence}\ 
M.Aguilar-Benitez\r\tute\madrid\ 
J.Alcaraz\r\tute{\madrid}\ 
G.Alemanni\r\tute\lausanne\
J.Allaby\r\tute\cern\
A.Aloisio\r\tute\naples\ 
M.G.Alviggi\r\tute\naples\
H.Anderhub\r\tute\eth\ 
V.P.Andreev\r\tute{\lsu,\peters}\
% ...
\newpage
%\rule{\textwidth}{0.4pt}
\begin{list}{A}{\itemsep=0pt plus 0pt minus 0pt\parsep=0pt plus 0pt minus 0pt
                \topsep=0pt plus 0pt minus 0pt}
\item[\aachen]
 III. Physikalisches Institut, RWTH, D-52056 Aachen, Germany$^{\S}$
\item[\nikhef] National Institute for High Energy Physics, NIKHEF, 
     and University of Amsterdam, NL-1009 DB Amsterdam, The Netherlands
\item[\mich] University of Michigan, Ann Arbor, MI 48109, USA
\item[\lapp] Laboratoire d'Annecy-le-Vieux de Physique des Particules, 
     LAPP,IN2P3-CNRS, BP 110, F-74941 Annecy-le-Vieux CEDEX, France
\item[\basel] Institute of Physics, University of Basel, CH-4056 Basel,
     Switzerland
\item[\lsu] Louisiana State University, Baton Rouge, LA 70803, USA
\item[\beijing] Institute of High Energy Physics, IHEP, 
  100039 Beijing, China$^{\triangle}$ 
\item[\bologna] University of Bologna and INFN-Sezione di Bologna, 
     I-40126 Bologna, Italy
\item[\tata] Tata Institute of Fundamental Research, Mumbai (Bombay) 400 005, India
\item[\ne] Northeastern University, Boston, MA 02115, USA
\item[\bucharest] Institute of Atomic Physics and University of Bucharest,
     R-76900 Bucharest, Romania
\item[\budapest] Central Research Institute for Physics of the 
     Hungarian Academy of Sciences, H-1525 Budapest 114, Hungary$^{\ddag}$
\item[\mit] Massachusetts Institute of Technology, Cambridge, MA 02139, USA
\item[\panjab] Panjab University, Chandigarh 160 014, India.
\item[\debrecen] KLTE-ATOMKI, H-4010 Debrecen, Hungary$^\P$
\item[\dublin] Department of Experimental Physics,
  University College Dublin, Belfield, Dublin 4, Ireland
\item[\florence] INFN Sezione di Firenze and University of Florence, 
     I-50125 Florence, Italy
\item[\cern] European Laboratory for Particle Physics, CERN, 
     CH-1211 Geneva 23, Switzerland
\item[\wl] World Laboratory, FBLJA  Project, CH-1211 Geneva 23, Switzerland
\item[\geneva] University of Geneva, CH-1211 Geneva 4, Switzerland
\item[\hefei] Chinese University of Science and Technology, USTC,
      Hefei, Anhui 230 029, China$^{\triangle}$
\item[\lausanne] University of Lausanne, CH-1015 Lausanne, Switzerland
\item[\lyon] Institut de Physique Nucl\'eaire de Lyon, 
     IN2P3-CNRS,Universit\'e Claude Bernard, 
     F-69622 Villeurbanne, France
\item[\madrid] Centro de Investigaciones Energ{\'e}ticas, 
     Medioambientales y Tecnol\'ogicas, CIEMAT, E-28040 Madrid,
     Spain${\flat}$ 
\item[\florida] Florida Institute of Technology, Melbourne, FL 32901, USA
\item[\milan] INFN-Sezione di Milano, I-20133 Milan, Italy
\item[\moscow] Institute of Theoretical and Experimental Physics, ITEP, 
     Moscow, Russia
\item[\naples] INFN-Sezione di Napoli and University of Naples, 
     I-80125 Naples, Italy
\item[\cyprus] Department of Physics, University of Cyprus,
     Nicosia, Cyprus
\item[\nymegen] University of Nijmegen and NIKHEF, 
     NL-6525 ED Nijmegen, The Netherlands
\item[\caltech] California Institute of Technology, Pasadena, CA 91125, USA
\item[\perugia] INFN-Sezione di Perugia and Universit\`a Degli 
     Studi di Perugia, I-06100 Perugia, Italy   
\item[\peters] Nuclear Physics Institute, St. Petersburg, Russia
\item[\cmu] Carnegie Mellon University, Pittsburgh, PA 15213, USA
\item[\potenza] INFN-Sezione di Napoli and University of Potenza, 
     I-85100 Potenza, Italy
\item[\prince] Princeton University, Princeton, NJ 08544, USA
\item[\riverside] University of Californa, Riverside, CA 92521, USA
\item[\rome] INFN-Sezione di Roma and University of Rome, ``La Sapienza",
     I-00185 Rome, Italy
\item[\salerno] University and INFN, Salerno, I-84100 Salerno, Italy
\item[\ucsd] University of California, San Diego, CA 92093, USA
\item[\sofia] Bulgarian Academy of Sciences, Central Lab.~of 
     Mechatronics and Instrumentation, BU-1113 Sofia, Bulgaria
\item[\korea]  The Center for High Energy Physics, 
     Kyungpook National University, 702-701 Taegu, Republic of Korea
\item[\purdue] Purdue University, West Lafayette, IN 47907, USA
\item[\psinst] Paul Scherrer Institut, PSI, CH-5232 Villigen, Switzerland
\item[\zeuthen] DESY, D-15738 Zeuthen, Germany
\item[\eth] Eidgen\"ossische Technische Hochschule, ETH Z\"urich,
     CH-8093 Z\"urich, Switzerland
\item[\hamburg] University of Hamburg, D-22761 Hamburg, Germany
\item[\taiwan] National Central University, Chung-Li, Taiwan, China
\item[\tsinghua] Department of Physics, National Tsing Hua University,
      Taiwan, China
\item[\S]  Supported by the German Bundesministerium 
        f\"ur Bildung, Wissenschaft, Forschung und Technologie
\item[\ddag] Supported by the Hungarian OTKA fund under contract
numbers T019181, F023259 and T037350.
\item[\P] Also supported by the Hungarian OTKA fund under contract
  number T026178.
\item[$\flat$] Supported also by the Comisi\'on Interministerial de Ciencia y 
        Tecnolog{\'\i}a.
\item[$\sharp$] Also supported by CONICET and Universidad Nacional de La Plata,
        CC 67, 1900 La Plata, Argentina.
\item[$\triangle$] Supported by the National Natural Science
  Foundation of China.
\end{list}
}
\vfill

%%% Local Variables: 
%%% mode: latex
%%% TeX-master: t
%%% End:

\newpage
%%
%%%%%%%%%%%%%%%%%%%%%%%%%%%%%%
%  TABLES
%%%%%%%%%%%%%%%%%%%%%%%%%%%%%%%%
\begin{table}[htb]
\begin{center}
\begin{tabular}{|c|c|c|c|c|c|}
\hline
$\langle\rs\rangle$ (GeV)  &  $ {\cal{L}} (\rm{pb}^{-1})$ & $N_{\rm{events}}$ & $N_{\rm{MC}}$ & $\epsilon$ & $\pi$  \\
\hline
 \hline
188.6 & 176.7 & 208 & 226.0 & 14.2\% & 88\%\\
\hline
191.6 & \phantom{1}29.7 &  \phantom{1}38 &  \phantom{1}37.9 & 14.3\% & 90\%\\
\hline
195.5 & \phantom{1}83.7 & 104 & 101.0 & 13.4\% & 92\%\\
\hline
199.5 &  \phantom{1}84.3 &  \phantom{1}97 &  \phantom{1}91.9 & 12.2\% & 93\%\\
\hline
201.7 &  \phantom{1}35.5 &  \phantom{1}36 &  \phantom{1}37.2 & 11.3\% & 93\%\\
\hline
205.1 &  \phantom{1}77.8 &  \phantom{1}75 &  \phantom{1}74.8 & 10.3\% & 93\%\\
\hline
206.6 & 138.9 & 108 & 120.8 &  \phantom{0}8.9\% & 91\%\\
\hline
\hline
198.2 & 626.6 & 666 & 689.6 & 12.0\% & 91\%\\
\hline
\end{tabular}
\caption[]{Average centre-of-mass energies, integrated luminosities (${\cal{L}}$), number of selected
events ($N_{\rm{events}}$), number of expected events ($N_{\rm{MC}}$), selection efficiency ($\epsilon$)
and percentage of
correct jet pairing ($\pi$) for the particle flow analysis. The combined figures are given in
the last row.}
\label{tab:selflow}
\end{center}
\end{table}

\begin{table}[htb]
\begin{center}
\begin{tabular}{|c|c|c|}
\hline
$\langle\rs\rangle$ (GeV)  &  $ R_{\rm{N}}$ & $R_{\rm{E}}$  \\
\hline
 \hline
188.6 & 0.820 $\pm$ 0.037 & 0.610 $\pm$ 0.047\\
\hline
191.6 & 0.929 $\pm$ 0.093 & 0.822 $\pm$ 0.133 \\
\hline
195.5 & 0.948 $\pm$ 0.059 & 0.774 $\pm$ 0.077 \\
\hline
199.5 & 1.004 $\pm$ 0.067 & 0.871 $\pm$ 0.095 \\
\hline
201.7 & 0.770 $\pm$ 0.086 & 0.626 $\pm$ 0.130 \\
\hline
205.1 & 1.033 $\pm$ 0.083 & 0.756 $\pm$ 0.111 \\
\hline
206.6 & 0.958 $\pm$ 0.068 & 0.781 $\pm$ 0.096 \\
\hline
\end{tabular}
\caption[]{Measured $R_{\rm{N}}$ and $R_{\rm{E}}$ values as a function of energy with
their statistical uncertainties.}
\label{tab:result}
\end{center}
\end{table}

\begin{table}[htb]
\begin{center}
\begin{tabular}{|c|c|}
\hline
Source    & $\sigma_{R_{\rm{N}}}$  \\
\hline
 \hline
Energy flow objects  &    0.016\\
\hline
$\qq$  fragmentation  &     0.009\\
\hline
WW fragmentation   &   0.008 \\
\hline
BEC  &     0.003\\
\hline
 4-jet background rate &    0.004\\
\hline
ZZ background  &     0.002\\
\hline
\hline
Total  &  0.021\\
\hline
\end{tabular}
\caption[]{Contributions to the systematic uncertainties on  $R_{\rm{N}}$.}
\label{tab:syst}
\end{center}
\end{table}

%%%%%%%%%%%%%%%%%%%%%%%%%%%%%%%%%%%%%%%%%%%%%%%%%%%%%%%%%%%%%%%%%%%%%%%%%%%%%%
\begin{table}[htb]
\begin{center}
\begin{tabular}{|c|c|}
\hline
   & $R_{\rm{N}}$ \\
\hline
\hline
 Data  & 0.915 $\pm$  0.023 $\pm$ 0.021  \\
\hline
\hline
PYTHIA no-CR   & 0.918 $\pm$  0.003 \\
\hline
 SKI ($k_{\rm{I}}$=0.6)  & 0.896 $\pm$  0.003 \\
\hline
 SKI ($k_{\rm{I}}$=3.0)  & 0.843 $\pm$  0.003 \\
\hline
 SKI 100\% & 0.762 $\pm$  0.003  \\
\hline
 SKII   & 0.916 $\pm$  0.003  \\
\hline
 ARIADNE no-CR   & 0.929 $\pm$  0.003 \\
\hline
 AR2   & 0.919 $\pm$  0.003  \\
\hline
 HERWIG no-CR   & 0.948 $\pm$  0.005 \\
\hline
 HERWIG CR  & 0.946 $\pm$  0.005  \\
\hline
\end{tabular}
\caption[]{Measured value of $R_{\rm{N}}$ and model predictions.}
\label{tab:model}
\end{center}
\end{table}

\clearpage
%%%%%%%%%%%%%%%%%%%%%%%%%%%%%%
%  FIGURES
%%%%%%%%%%%%%%%%%%%%%%%%%%%%%%%%
%%%%%%%%%%%%%%%%%%%%%%%%%%%%%%%%%%%%%%%%%%%%%%%%%%%%%%%%%%%%%%%%%%%%%%%%%%%%%%

%%%%%%%%%%%%%%%%%%%%%%%%%%%%%%%%%%%%%%%%%%%%%%%%%%%%%%%%%%%%%%%%%%%%%%%%%%%%%%
%%%%%%%%%%%%%%%%%%%%%%%%%%%%%%%%%%%%%%%%%%%%%%%%%%%%%%%%%%%%%%%%%%%%%%%%%%%%%
%%%%%%%%%%%%%%%%%%%%%%%%%%%%%%%%%%%%%%%%%%%%%%%%%%%%%%%%%%%%%%%%%%%%%%%%%%%%%%
%%%%%%%%%%%%%%%%%%%%%%%%%%%%%%%%%%%%%%%%%%%%%%%%%%%%%%%%%%%%%%%%%%%%%%%%%%%%%
%%%%%%%%%%%%%%%%%%%%%%%%%%%%%%%%%%%%%%%%%%%%%%%%%%%%%%%%%%%%%%%%%%%%%%%%%%%%%%

%%%%%%%%%%%%%%%%%%%%%%%%%%%%%%%%%%%%%%%%%%%%%%%%%%%%%%%%%%%%%%%%%%%%%%%%%%%%%%

%%%%%%%%%%%%%%%%%%%%%%%%%%%%%%%%%%%%%%%%%%%%%%%%%%%%%%%%%%%%%%%%%%%%%%%%%%%%%%
%%%%%%%%%%%%%%%%%%%%%%%%%%%%%%%%%%%%%%%%%%%%%%%%%%%%%%%%%%%%%%%%%%%%%%%%%%%%%%
%%%%%%%%%%%%%%%%%%%%%%%%%%%%%%%%%%%%%%%%%%%%%%%%%%%%%%%%%%%%%%%%%%%%%%%%%%%%%
\end{document}